\documentclass[twocolumn,superscriptaddress,amsmath,amssymb,showpacs,pre]{revtex4-2}

\usepackage{graphicx}
\usepackage{color}
\renewcommand{\phi}{\varphi}

\begin{document}

\title{Prediction of crystal structures and motifs in the Fe-Mg-O system at Earth’s core pressures}

\author{Renhai Wang}
    \affiliation{Department of Physics, University of Science and Technology of China, Hefei 230026, China}
    \affiliation{Department of Physics, Iowa State University, Ames, Iowa 50011, USA}
\author{Yang Sun}
    \email[Email: ]{ys3339@columbia.edu}
    \affiliation{Department of Applied Physics and Applied Mathematics, Columbia University, New York, NY, 10027, USA}
\author{Renata M. Wentzcovitch}
\email[Email: ]{rmw2150@columbia.edu}
    \affiliation{Department of Applied Physics and Applied Mathematics, Columbia University, New York, NY, 10027, USA}
    \affiliation{Department of Earth and Environmental Sciences, Columbia University, New York, NY, 10027, USA}
    \affiliation{Lamont–Doherty Earth Observatory, Columbia University, Palisades, NY, 10964, USA}
\author{Feng Zheng}
    \affiliation{Department of Physics, Xiamen University, Xiamen 361005, China}
\author{Yimei Fang}
    \affiliation{Department of Physics, Xiamen University, Xiamen 361005, China}
\author{Shunqing Wu}
    \affiliation{Department of Physics, Xiamen University, Xiamen 361005, China}  
\author{Zijing Lin}
    \affiliation{Department of Physics, University of Science and Technology of China, Hefei 230026, China}
\author{Cai-Zhuang Wang}
    \affiliation{Department of Physics, Iowa State University, Ames, Iowa 50011, USA}
\author{Kai-Ming Ho}
    \affiliation{Department of Physics, Iowa State University, Ames, Iowa 50011, USA}

\date{Feb. 4, 2021}

\begin{abstract}

Fe, Mg, and O are among the most abundant elements in terrestrial planets. While the behavior of the Fe-O, Mg-O, and Fe-Mg binary systems under pressure have been investigated, there are still very few studies of the Fe-Mg-O ternary system at relevant Earth's core and super-Earth's mantle pressures. Here, we use the adaptive genetic algorithm (AGA) to study ternary Fe$_x$Mg$_y$O$_z$ phases in a wide range of stoichiometries at 200 GPa and 350 GPa. We discovered three dynamically stable phases with stoichiometries FeMg$_2$O$_4$, Fe$_2$MgO$_4$, and FeMg$_3$O$_4$ with lower enthalpy than any known combination of Fe-Mg-O high-pressure compounds at 350 GPa. With the discovery of these phases, we construct the Fe-Mg-O ternary convex hull. We further clarify the composition- and pressure-dependence of structural motifs with the analysis of the AGA-found stable and metastable structures. Analysis of binary and ternary stable phases suggest that O, Mg, or both could stabilize a BCC iron alloy at inner core pressures.

\end{abstract}

\maketitle

\section{Introduction}
O, Fe, Si, Mg, Al, and Ca (CMAS+F) are the most abundant elements in terrestrial planets \cite{1}. Among these planets, Earth provides essential general information, yet it is incompletely deciphered. All CMAS+F elements are lithophile (rock-loving) elements and are present in the Earth's rocky mantle and crust. Fe is the predominant element in the core and is a siderophile (metal-loving) element as well. Based on current knowledge, this classification is believed to be valid in the pressure and temperature (PT) range achieved in Earth's interior. Seismology and high-pressure data on iron shows that the Earth's core is $\sim$ 5-10 wt\% \cite{2} less dense than iron at expected conditions, i.e., $\sim$ 136-364 GPa and $\sim$ 4,000-6,500 K \cite{3,4,5,6,7,8}. This indicates the presence of lighter elements in the core partitioned differently between its solid and liquid regions \cite{9}. Extensive research has been carried out experimentally and computationally to shed light on the light elements' nature. Despite much progress, there is still a great deal of uncertainty \cite{2,10,11,12}. With every experimental or computational development, this question is revisited from a different angle. In particular, the development of materials discovery methods \cite{13,14,15,16,17,18,19,20} has propelled the exploration of novel chemistries under pressure, which has fueled the debate on the possible nature of light elements in the core.  
Among the CMAS elements, the most likely light element candidates in the core are Si, and O. O is considered a required element today, a view that has evolved within the last decade \cite{21,22,23,24}. The volatile elements S, C, and H are also regarded as likely candidates, but the abundances of C and H on Earth are still largely unconstrained. Mg, Al, and Ca have been branded lithophile elements up to core pressures. But the recent computational discovery of Mg-Fe compounds up to inner core pressures \cite{25} suggests the possibility of Mg turning siderophile and its presence in the core. The formation of Fe-O \cite{16} and Fe-Si \cite{15} compounds with variable stoichiometry have been investigated using materials discovery methods. The theoretical prediction of pyrite-type FeO$_2$ \cite{16} and its experimental confirmation \cite{26,27} has been one of the greatest successes of this approach, which rarely explores the possibility of ternary compounds \cite{17}. Given the present understanding that O is a required element in the outer core and should also exist in the inner core, we explore the possible formation of Fe-Mg-O compounds at typical core pressures of $\sim$ 350 GPa. The investigation of solid compounds provides the most critical information. Light elements must be present in both solid and liquid phases but more abundantly in the liquid phase. It is energetically more costly to accommodate these elements in the solid phase, a geochemical definition of incompatible elements. Therefore, the discovery of thermodynamically stable Fe-Mg-O solids is essential to investigate Mg's presence in the core. 

Besides being essential for addressing Mg's presence in the Earth's core, the present study of Fe-Mg-O solids has significant ramifications for the mantle of terrestrial planets larger than Earth whose interiors can reach much higher pressures and temperatures. The B1-type isomorphous alloy (Mg$_{1-x}$Fe$_x$)O, ferropericlase (x $<$ 0.5) or magnesiumwüstite (x $>$ 0.5) is the second most abundant phase of the Earth's lower mantle. Ferropericlase is the thermodynamically stable form of the Fe-Mg-O ternary compound up to $\sim$ 135 GPa and $\sim$ 4,000 K, i.e., CMB conditions in the Earth. The higher pressures and temperatures expected in Super-Earths, e.g., $\sim$ 4 TPa and $\sim$ 9,000 K at the CMB in a 20 M$_{\bigoplus}$ terrestrial planet \cite{28}, raises the possibility of other compounds and alloy structures with other compositions. The existence of other stable ternary phases under pressure may induce decomposition and recombination reactions between Fe-Mg-O, Fe-O, Mg-O, and Fe-Mg compounds under pressure, similar to what has been observed in the Si-Mg-O system \cite{17,18}. Therefore, the current research can also provide the first glimpses on the essential (Mg$_{1-x}$Fe$_x$)O alloy behavior in these planets.

This paper is organized as follows. In the next section, we describe the computational method used in this work. Sec. III presents the crystal structure search results, the ternary convex hull, and the predominant structural motifs in low enthalpy structures. Section IV discusses some potential geophysical implications of these results. Conclusions are present in Section V.

\section{Methods}
Crystal structures of Fe-Mg-O at high pressure were investigated using the adaptive genetic algorithm (AGA)\cite{17,29}. This method integrates auxiliary interatomic potentials and \textit{ab initio} calculations adaptively. The auxiliary interatomic potentials accelerate crystal structure searches in the genetic algorithm (GA) loop. At the same time, \textit{ab initio} calculations are used to adapt the potentials after several GA generations to ensure accuracy. The structure searches were only constrained by the chemical composition, without any assumption on the Bravais lattice type, symmetry, atomic basis, or unit cell dimensions. In our AGA searches, the enthalpy was used as the selection criteria for optimizing the candidate pool. The candidate structure pool size in GA search is 128. At each GA generation, 32 new structures are generated from the parent structure pool via a mating procedure described in \cite{30}. The structures in the pool were updated by keeping the lowest-energy 128 structures. The structure search with a given auxiliary interatomic potential sustained 600 consecutive GA generations. Then, 16 structures from the GA search were randomly selected for \textit{ab initio} calculations to re-adjust the interatomic potential parameters for the next round of the GA search. This sequence of steps was repeated 40 times. 
In the AGA search in the Fe-Mg-O system, interatomic potentials based on the embedded-atom method (EAM) \cite{31} were chosen as the auxiliary classical potential. In EAM, the total energy of an N-atom system is described by 
\begin{equation}
E_{total}=\frac{1}{2}\sum_{i,j\left(i\neq j\right)}^{N}{\phi\left(r_{ij}\right)+\sum_{i}{F_i\left(n_i\right)}}
\label{eq1},
\end{equation}
where $\phi\left(r_{ij}\right)$ is the pair term for atoms $i$ and $j$ at a distance $r_{ij}$. $F_i\left(n_i\right)$ is the embedded term with electron density term $n_i=\sum_{j\neq i}{\rho_j\left(r_{ij}\right)}$ at the site occupied by atom $i$. The fitting parameters in the EAM formula are chosen as follows: The parameters for Fe-Fe and Mg-Mg interactions were taken from the literature \cite{32}. Other pair interactions (O-O, Fe-Mg, Fe-O and Mg-O) were modeled with the Morse function,
\begin{equation}
\phi\left(r_{ij}\right)=D\left[e^{-2\alpha\left(r_{ij}-r_0\right)}-2e^{-\alpha\left(r_{ij}-r_0\right)}\right]
\label{eq2},
\end{equation}
where $D$, $\alpha$, $r_0$ are fitting parameters. The density function for O atoms are modeled by an exponentially decaying function,

\begin{equation}
\rho\left(r_{ij}\right)=\alpha e\ x\ p{\left[-\beta\left(r_{ij}-r_0\right)\right]}
\label{eq3},
\end{equation}
where $\alpha$ and $\beta$ are fitting parameters. The form proposed by Benerjea and Smith \cite{33} was used as the embedding function with fitting parameters $F_0$, $\gamma$ as
\begin{equation}
F\left(n\right)=F_0\left[1-\gamma\ In\ n\right]n^\gamma
\label{eq4}.
\end{equation}

For Fe and Mg, the parameters of the density function and embedding function were taken from ref.\cite{32} as well. In the AGA scheme\cite{17}, the potential parameters were adjusted adaptively by fitting to the \textit{ab initio} energies, forces, and stresses of selected structures. The fitting process was performed using the force-matching method with a stochastic simulated annealing algorithm implemented in the POTFIT code\cite{34,35}. 

\textit{Ab initio} calculations were carried out using the projector augmented wave (PAW) method\cite{36} within density functional theory (DFT) as implemented in the VASP code \cite{37,38}. The exchange and correlation energy are treated without the spin-polarized generalized gradient approximation (GGA) and parameterized by the Perdew-Burke-Ernzerhof formula (PBE) \cite{39}. A plane-wave basis was used with a kinetic energy cutoff of 520 eV, and the convergence criterion for the total energy was set to $10^{-5}$ eV. Monkhorst-Pack's sampling scheme \cite{40} was adopted for Brillouin zone sampling with a k-point grid of $2\pi \times 0.033 \text{\AA}^{-1}$, and the unit cell lattice vectors (both the unit cell shape and size) are fully relaxed under fixed pressure (200 GPa and 350 GPa) together with the atomic coordinates until the force on each atom is less than 0.01 eV/$\text{\AA}$. The phonon dispersions were computed with density functional perturbation theory (DFPT) implemented in the VASP code and the Phonopy software \cite{41}.
The formation enthalpy (H$_f$) of compound Fe$_x$Mg$_y$O$_z$ was calculated as
\begin{equation}
H_f=\frac{H\left({\rm Fe}_x{\rm Mg}_yO_z\right)-xH\left(Fe\right)-yH\left(Mg\right)-zH\left(O\right)}{x+y+z}
\label{eq5},
\end{equation}
where $H\left({\rm Fe}_x{\rm Mg}_yO_z\right)$ is the total enthalpy of the Fe$_x$Mg$_y$O$_z$ alloy. $H\left(Fe\right)$, $H\left(Mg\right)$ and $H\left(O\right)$ are the enthalpy of the ground state of Fe, Mg, and O at corresponding pressures, i.e., hcp-Fe, bcc-Mg, and $\zeta$-O$_2$, respectively.

\section{Results and Discussion}
\subsection{AGA search for ternary Fe-Mg-O phases}
\begin{figure}
\includegraphics[width=0.75\textwidth]{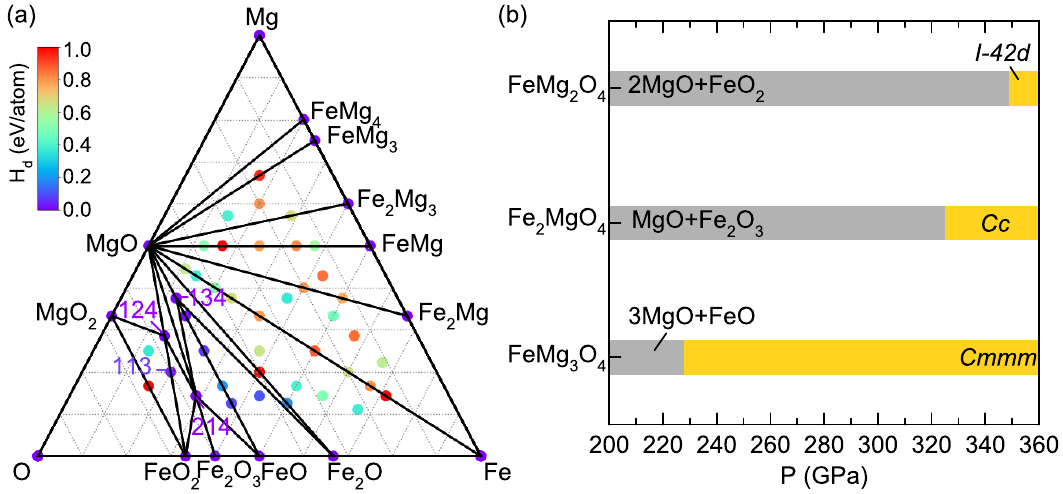}
\caption{\label{fig:fig1} (a) AGA search results for the Fe-Mg-O system at 350 GPa. The dots represent searched ternary compositions. The color bar corresponds to the relative enthalpy above the convex hull. The ground-states ($H_d=0$) on the convex hull are connected. The new phases are indicated by the text. 124, 134, 214 and 113 represents FeMg$_2$O$_4$, FeMg$_3$O$_4$, Fe$_2$MgO$_4$ and FeMgO$_3$, respectively. (b) Stability range of discovered ternary ground states at $T=0K$. The gray bars indicates the pressure range of decomposition.}
\end{figure}

In Fig.~\ref{fig:fig1}(a) we present the AGA results of Fe-Mg-O at 350 GPa. For the sake of simplicity, all chemical formulae are expressed as Fe/Mg/O reduced ratios. For example, 123 represents the compound with FeMg$_2$O$_3$. During the structural search, we select a range of different stoichiometries surrounding 111 composition (i.e., 211, 121, 112, 311, 131, 113, 411, 141, 114, 221, 122, 212, 331, 133, 313, 441, 144, 414, 332, 233, 323, 321, 312, 123, 132, 231, 213, 421, 412, 124, 142, 241, 214, 431, 413, 134, 143, 341 and 314) with 2 or 4 formula units to perform the AGA search (up to 32 atoms per primitive cell). After the AGA search, we use the following method to determine the stability of compounds. For a ternary compound $\text{A}_x\text{B}_y\text{C}_z$, we select three existing compounds $\text{A}_{x1}\text{B}_{y1}\text{C}_{z1}$, $\text{A}_{x2}\text{B}_{y2}\text{C}_{z2}$, and $\text{A}_{x3}\text{B}_{y3}\text{C}_{z3}$ on the diagram; these can also be elementary or binary end-members. If $\text{A}_x\text{B}_y\text{C}_z$ can be written as $a\ \times\ \text{A}_{x1}\text{B}_{y1}\text{C}_{z1} + b\ \times\ \text{A}_{x2}\text{B}_{y2}\text{C}_{z2} + c\ \times\ \text{A}_{x3}\text{B}_{y3}\text{C}_{z3}$ with $a \geq 0$, $b \geq 0$ and $c \geq 0$, we compute its relative enthalpy $\Delta H = H(A_xB_yC_z) – a \times H(A_{x1}B_{y1}C_{z1})–b\times H(A_{x2}B_{y2}C_{z2}) – c \times H(A_{x3}B_{y3}C_{z3})$. If $\Delta H \le 0$ for all scanned combinations of existing phases on the diagram, $A_xB_yC_z$ is determined as an energetic ground state. The energetic ground states form the convex hull as shown in Fig. 1(a). $H_d$ is introduced as the enthalpy above the convex hull to represent the relative stability on the phase diagram. By definition, all the ground-state phases have $H_d=0$.

The AGA search found three new ternary ground state compounds at 350 GPa: FeMg$_2$O$_4$, Fe$_2$MgO$_4$, and FeMg$_3$O$_4$. They define the current Fe-Mg-O ternary phase diagram. The stability of these phases from 200 GPa and 350 GPa is shown in Fig.~\ref{fig:fig1}(b). This stability pressure range is computed by considering the relative stability of these phases against decomposition into all end-members (see Supplementary Fig. 1 for the stability range of all ground-state phases). We will discuss the construction of the phase diagram in the next section. We also identify low-enthalpy metastable structures such as FeMgO$_3$ with enthalpy very close to the convex hull ($H_d=18$ meV/atom). Here we first analyze these new ground states and low-enthalpy structures.

\onecolumngrid

\twocolumngrid

\begin{figure}
\includegraphics[width=0.48\textwidth]{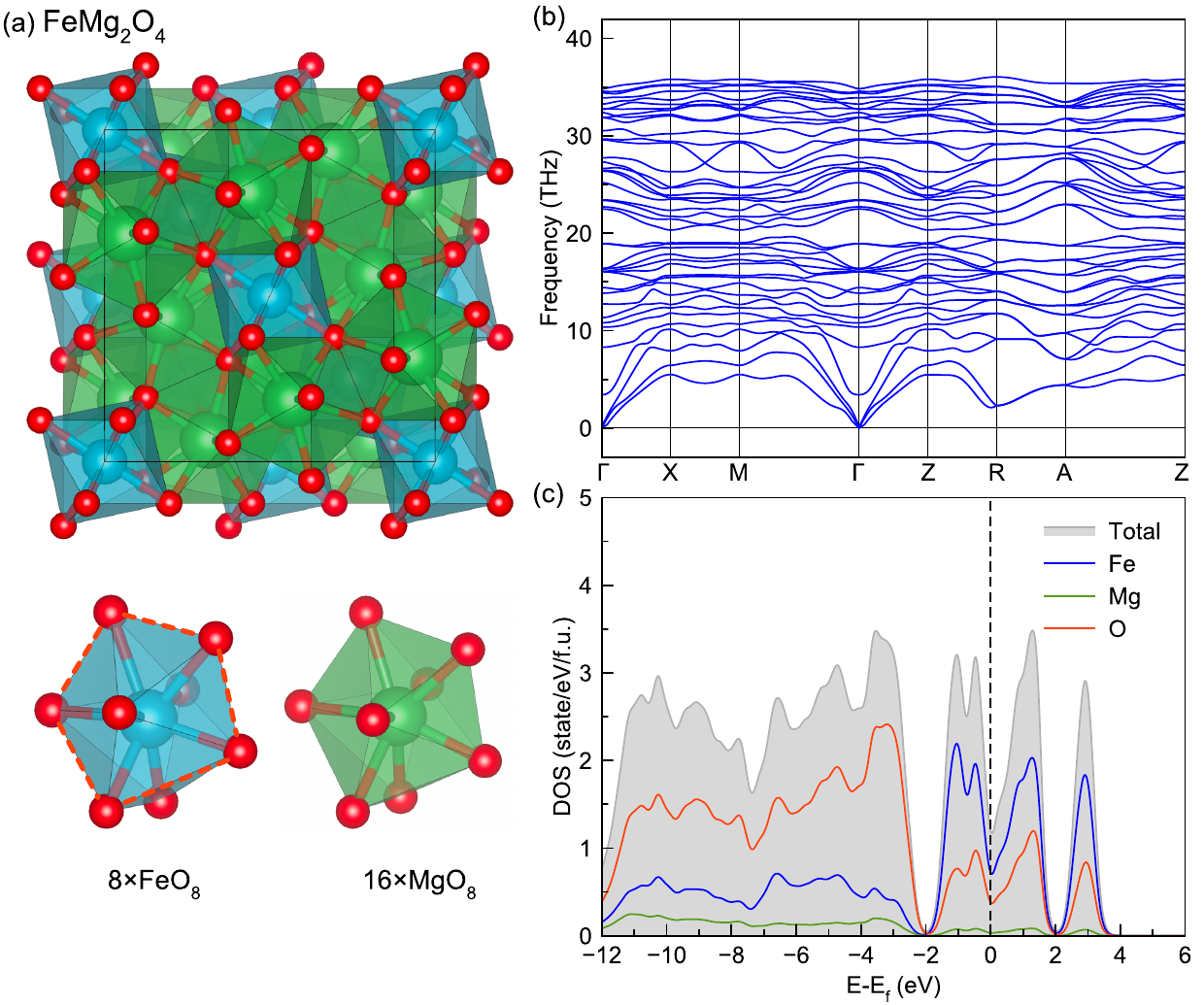}
\caption{\label{fig:fig2} (a) Atomic structure of \textit{I-42d} FeMg$_2$O$_4$ and Fe and Mg coordination polyhedra. Blue is Fe, green is Mg and red is O. Red dashed lines indicate a pentagonal cap; (b) phonon dispersion; (c) electronic density of states.}
\end{figure}

Figure~\ref{fig:fig2} shows the atomic structure, phonon dispersion, and electronic density of states for tetragonal FeMg$_2$O$_4$ with space group \textit{I-42d}. Fe and Mg atoms are coordinated with eight oxygen atoms to form a similar MO$_8$ (M for metal) polyhedra. Unlike a typical cubic polyhedron, this MO$_8$ consists only of triangular faces. These triangular faces form pentagonal caps and are similar to the Frank-Kasper polyhedra \cite{42}. The Fe- and Mg-centered MO8 polyhedra pack in various edge- and face-sharing arrangements. This structure is the same as the \textit{I-42d}-type Mg$_2$SiO$_4$ found previously \cite{17,43}. As shown in Fig.~\ref{fig:fig1}(b), this phase becomes the ground state at 349 GPa. Below this pressure, it decomposes into MgO and FeO$_2$. As shown in Fig.~\ref{fig:fig2}(b), there are no imaginary phonon mode frequencies, which confirms this phase is dynamically stable. The electronic density of state in Fig.~\ref{fig:fig2}(c) shows a metallic state with somewhat localized Fe and O states near the Fermi level.

\begin{figure}
\includegraphics[width=0.48\textwidth]{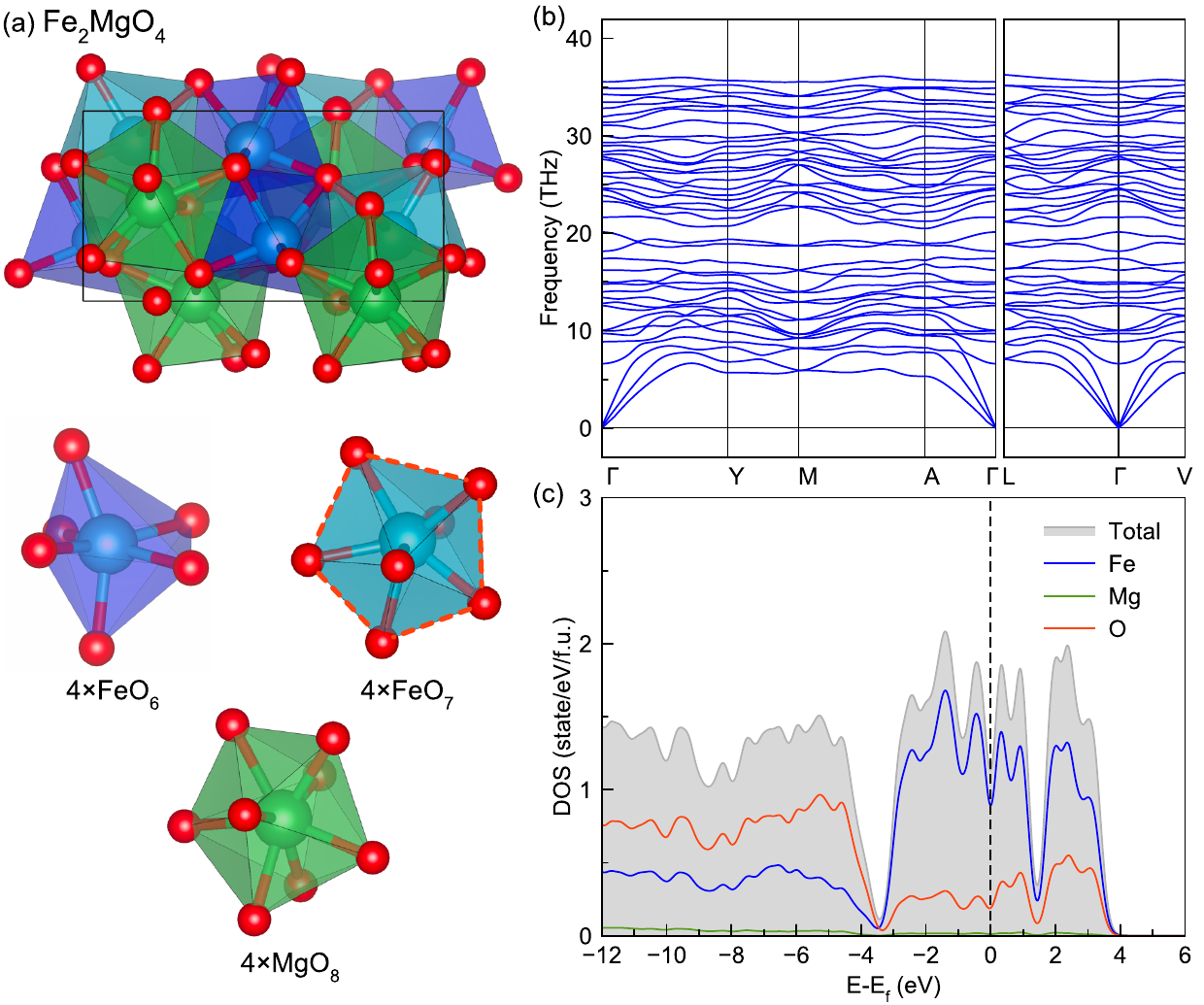}
\caption{\label{fig:fig3} (a) Atomic structure \textit{Cc} Fe$_2$MgO$_4$ and Fe and Mg coordination polyhedra. Blue is Fe, green is Mg and red is O. Red dashed lines indicate a pentagonal cap; (b) phonon dispersion; (c) electronic density of states.}
\end{figure}

Figure~\ref{fig:fig3} shows the atomic structure, phonon dispersion, and electronic density of states for monoclinic Fe$_2$MgO$_4$ with space group \textit{Cc}. By inspecting the structure, we identify the same type of MgO$_8$ polyhedra found in \textit{I-42d} FeMg$_2$O$_4$. However, the Fe-O polyhedra are more complicated. It contains two polyhedral types, one six-fold and one seven-fold coordinated as shown in Fig.~\ref{fig:fig3}(a). The FeO$_6$ is a highly distorted octahedron. The FeO$_7$ also shows the pentagonal cap similar to the MO$_8$ polyhedron found in \textit{I-42d} FeMg$_2$O$_4$. As shown in Fig.~\ref{fig:fig1}(b), this phase's stability pressure range against the decomposition into MgO and Fe$_2$O$_3$ starts at 325GPa. Phonon calculations confirm its dynamic stability. The electronic density of states also indicates this phase is metallic in Fig.~\ref{fig:fig3}(c).

\begin{figure}
\includegraphics[width=0.48\textwidth]{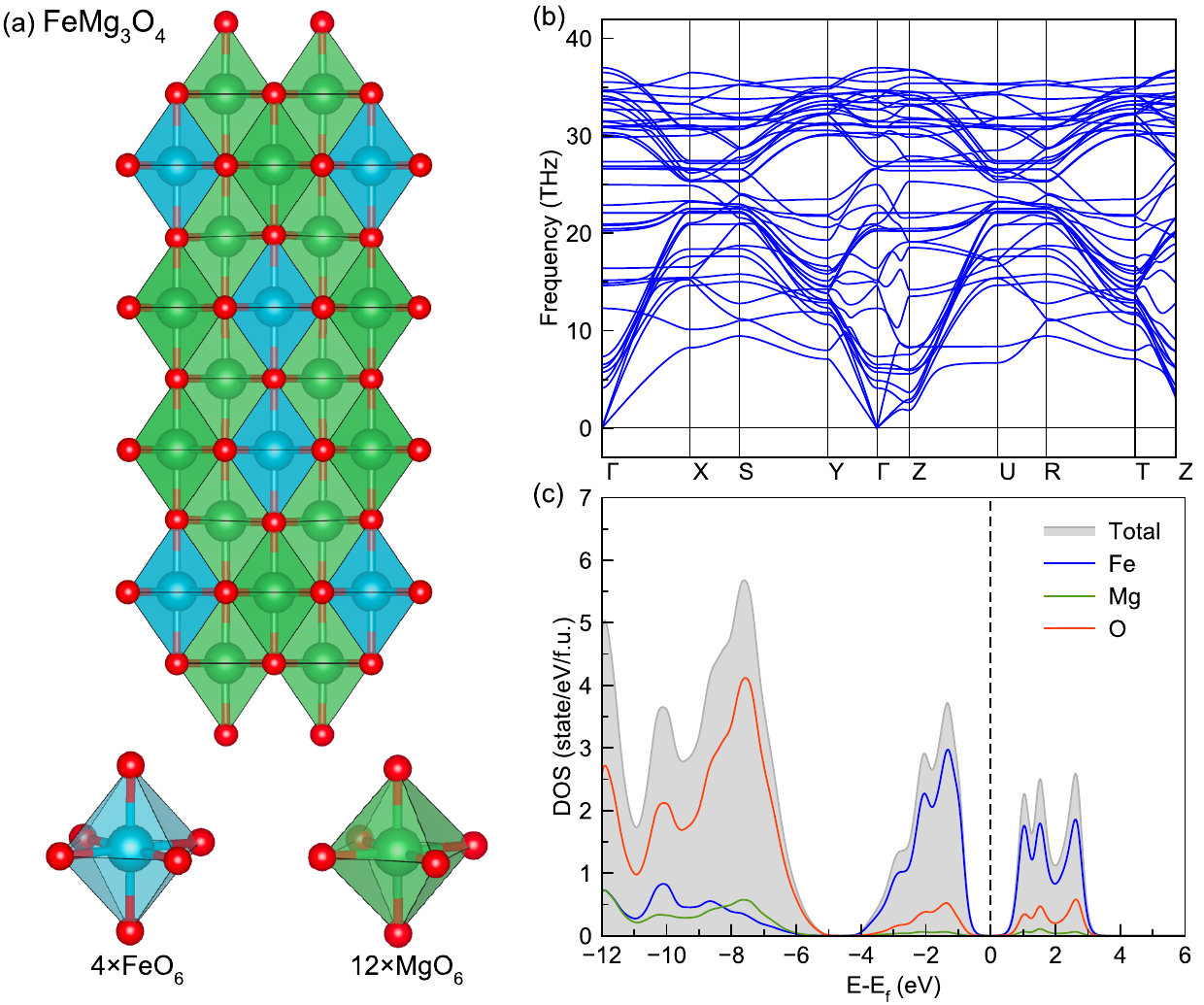}
\caption{\label{fig:fig4} (a) Atomic structure of \textit{Cmmm} FeMg$_3$O$_4$ and Fe and Mg coordination polyhedral. Blue is Fe, green is Mg, and red is O; (b) phonon dispersion; (c) electronic density of states.}
\end{figure}

The FeMg$_3$O$_4$ \textit{Cmmm} structure in Fig.~\ref{fig:fig4} shows Fe-O and Mg-O octahedral building blocks. Visually it is similar to ferropericlase (Fe$_{1-x}$Mg$_x$)O, which has a NaCl-type (B1) structure. However, unlike the cubic structure and random Fe/Mg cation site occupancies of ferropericlase, FeMg$_3$O$_4$ is an orthorhombic structure with ordered Fe/Mg site occupancies. The octahedra in the FeMg$_3$O$_4$ structure are Jahn-Teller distorted because of the orthorhombic symmetry. This phase becomes stable against decomposition into FeO and MgO at $\sim$ 228 GPa. Phonon calculations in Fig.~\ref{fig:fig4}(b) also confirm its dynamical stability. Unlike the metallic Fe$_2$MgO$_4$ and FeMg$_2$O$_4$ phases, FeMg$_3$O$_4$ \textit{Cmmm} is a semiconductor. 

\begin{figure}
\includegraphics[width=0.48\textwidth]{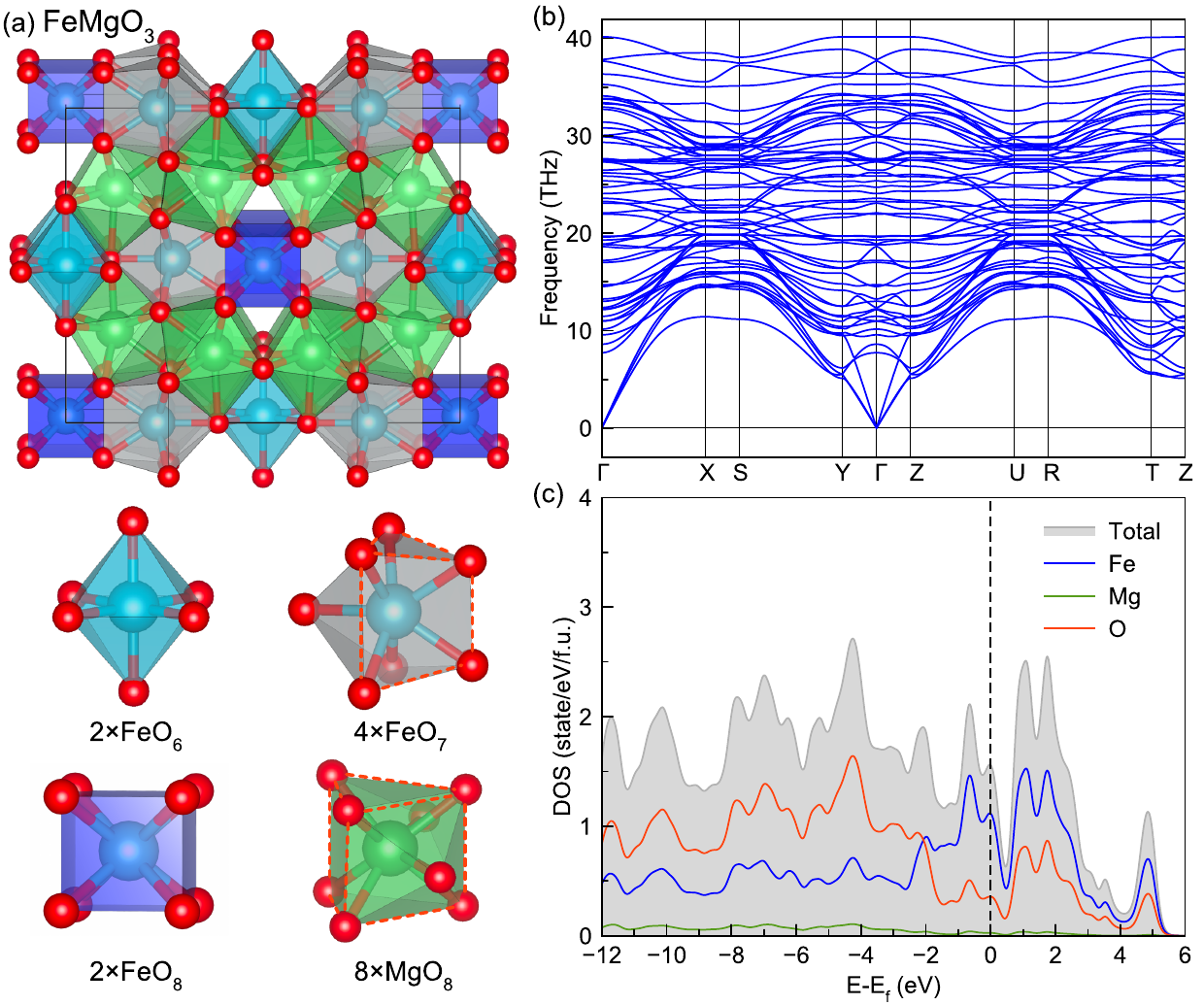}
\caption{\label{fig:fig5} (a) Atomic structure of metastable \textit{Immm} FeMgO$_3$ and Fe and Mg coordination polyhedra. Blue is Fe, green is Mg and red is O. Red dashed lines indicate the trigonal prism; (b) phonon dispersion; (c) electronic density of states.}
\end{figure}

Besides the ternary ground states, we also analyze a metastable FeMgO$_3$ \textit{Immm} structure with $H_d=18$ meV/atom. This enthalpy difference is so small that the compound may become stable at high temperatures. Structural analysis of FeMgO$_3$ \textit{Immm} in Fig.~\ref{fig:fig5} shows an interesting combination of various polyhedra. Iron shows three different oxygen coordination pulyhedra FeO$_6$, FeO$_7$ and FeO$_8$. The FeO$_6$ is an octahedron. The FeO$_7$ is a trigonal prism with an extra rectangular face capping neighbor. The FeO$_8$ is a cube. MgO shows one coordination polyhedron type, MgO$_8$, not a cube but a triangular prism with two rectangular face capping oxygens. Such a combination of octahedra and prisms is similar to the building blocks in the complex Fe$_2$O$_3$ polytypes \cite{44}. The appearance of cubic polyhedron is consistent with the observation of the CsCl-type (B2) structure of FeO at Earth's core conditions \cite{11}. This FeMgO$_3$ \textit{Immm} structure shows an intermediate packing between NaCl-type FeO/MgO phase to the CsCl-type FeO/MgO phases. This phase is dynamically stable and metallic, as shown by the phonon dispersion and the electronic density of states in Fig.~\ref{fig:fig5}.

\subsection{Construction of the Fe-Mg-O ternary convex hull}
In this section, we discuss the construction of the ternary phase diagram and convex hull. In a binary system, the compositional space is one-dimensional so that the convex hull is a curve connecting the formation enthalpies of ground-state phases. In a ternary system, the compositional space is two-dimensional, and the convex hull consists of surface segments connecting the formation enthalpies of three stable phases, as shown in Fig.~\ref{fig:fig1}. In a discrete compositional space, these surface segments are triangles. Any new structure having formation enthalpy below this convex hull surface will be a new ground state. The convex hull surface needs to be reconstructed after the discovery of any new stable phase.

\begin{figure}[b]
\includegraphics[width=0.63\textwidth]{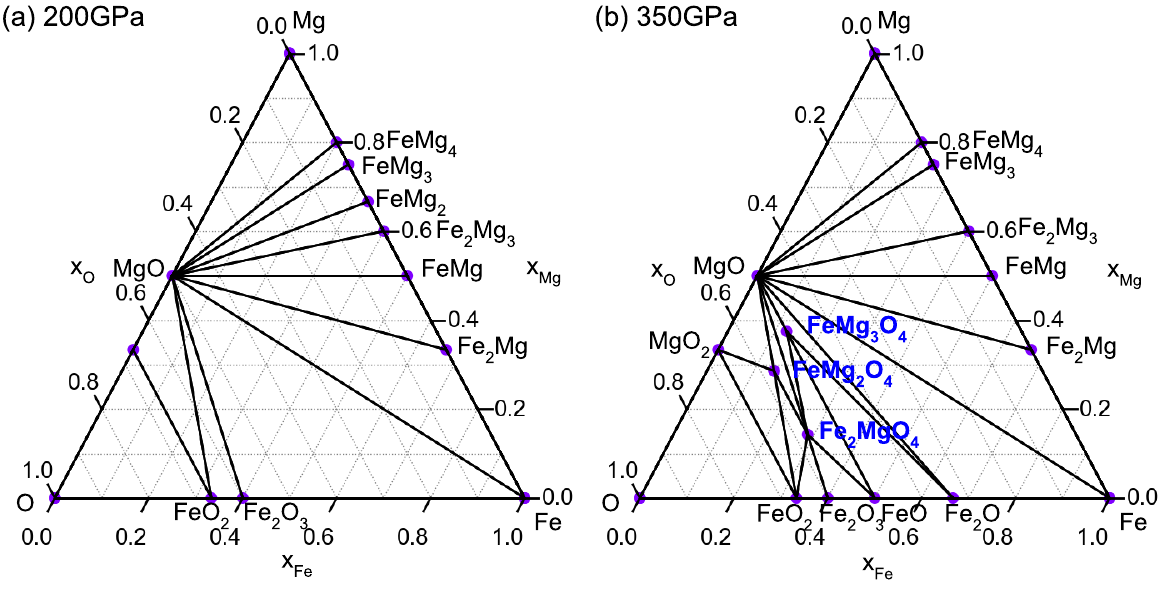}
\caption{\label{fig:fig6} Ground-state Fe-Mg-O phases and convex hull at (a) 200 GPa and (b) 350 GPa. The blue text refers to stable ternary phases.}
\end{figure}

For binary references at 350 GPa, the ground-state phases of Fe-O, Fe-Mg, and Mg-O have been investigated in Refs. \cite{16}, \cite{25} and \cite{45}, respectively (see Supplementary Fig. 2 for their crystal structures). Because these crystal structure searches have already covered the current study's pressure range, we do not perform new AGA searches for the binary phases. Still, we re-calculate the energetics of the previously found crystal structures. \textit{Ab initio} calculations confirm these reported relative phase stabilities of Fe-O \cite{16}, Fe-Mg \cite{25}, and Mg-O \cite{45}. By exploring an experimental database \cite{46}, we find two previously reported MgFe$_2$O$_4$ stoichiometric compounds [47,48]. However, our calculations indicate these phases are metastable at Earth's core pressures. Based on these ground-state binary phases, we established the Fe-Mg-O ternary system's convex-hull shown in Fig.~\ref{fig:fig6}. At 200 GPa, all the AGA searched ternary compounds have relatively higher enthalpy than the elementary or binary ground-state references. Therefore no Fe-Mg-O stoichiometric phase can be a ternary ground state at 200 GPa. At 350 GPa three ternary phases become stable ground states. Detailed energetics and crystallographic information on these ground-state phases is given in Supplementary Tables S1, Table S2 and Table S3.

\onecolumngrid

\clearpage

\twocolumngrid

\subsection{Analysis of structural motifs under pressure}
\begin{figure}
\includegraphics[width=0.8\textwidth]{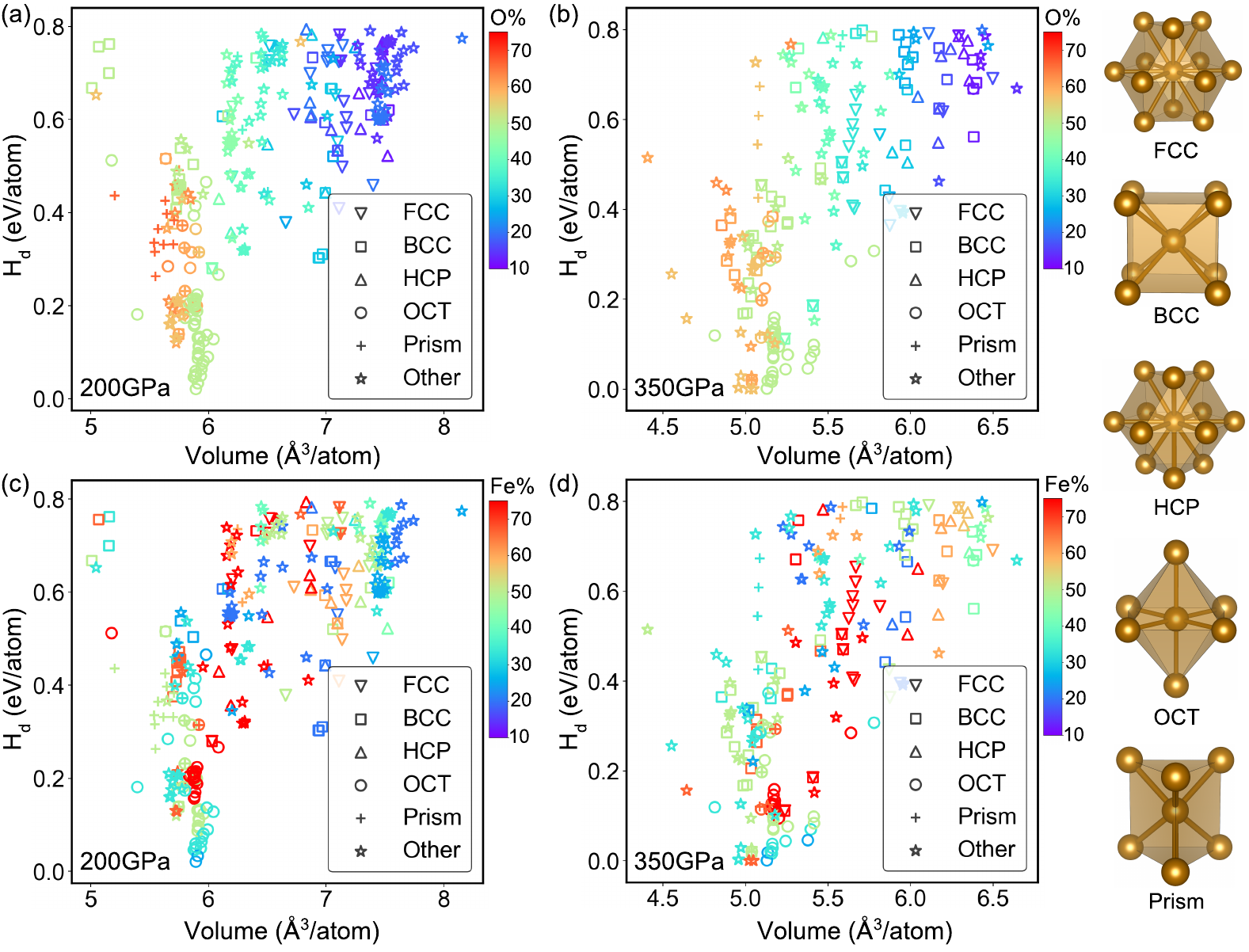}
\caption{\label{fig:fig7} Scatter plot of enthalpies above convex-hull (Hd) and volume for both stable and metastable Fe-Mg-O structures from AGA search. The color bar in (a) and (b) represent total oxygen concentration as $O\%=\frac{n(O)}{n(Fe)+n(Mg)+n(O)}\times100\%$. The color bar in (c) and (d) represent iron concentration in Fe and Mg as $Fe\%=\frac{n(Fe)}{n(Fe)+n(Mg)}\times100\%$. The right panel shows the template motifs.}
\end{figure}

Since Fig.~\ref{fig:fig6} suggests a strong effect of pressure on these phases' stability, we now investigate how the structural motifs change under pressure. Because the current calculation does not include temperature effects on phase relations and at finite temperatures, the ground states may differ. We now focus on the ground-state phases and metastable phases with formation enthalpy within 0.8 eV/atom ($\sim$9,000 K) above the convex hull. Phases in this energy range provide much better statistical information than just ground states. Here we employ a cluster alignment (CA) method \cite{49,50,51} to analyze the motifs of Fe-centered and Mg-centered clusters in each structure. The clusters here are defined by a center atom and its first-shell neighbor atoms. The cluster alignment method identifies the similarity between an as-extracted cluster and the perfect template cluster. Here we use typical motifs in the metallic alloys, including FCC-, BCC- and HCP-type clusters as the templates. We also include common motifs in Fe-O binary compounds which are OCT (octahedron) and Prism (trigonal prism) \cite{44}. The snapshots of these motifs are shown in Fig.~\ref{fig:fig7}. We first align the atomic cluster against these five motifs to check its similarity with the CA method. If the cluster does not match any of the above five polyhedra, it is marked as an "other" type. In Fig.~\ref{fig:fig7} we show scatter-type plots of enthalpy differences between these structures and that of the convex-hull and their respective atomic volumes. We differentiate the cluster type and O or Fe concentrations with symbols and colors, respectively. Since we obtain similar results between the analysis of Fe-centered clusters and Mg-centered clusters, we only show the analysis of Fe-centered clusters in Fig.~\ref{fig:fig7} and provide the analysis of Mg-centered clusters in Supplementary Fig. S3.

\onecolumngrid

\twocolumngrid

At 200 GPa, Figs.~\ref{fig:fig7}(a,c) indicate that crystal structures with octahedral type clusters and 50 mole\% oxygen concentration generally have lower energy than the other structures. While there is no ternary ground state at 200 GPa, these structures are concentrated within an energy range very close to the convex hull. Therefore, it is likely that they become ground states at high temperatures. These structures are all similar to the NaCl-type B1 structure and have variable Fe/Mg ratios and occupancies, which is essentially (Mg$_{1-x}$Fe$_x$)O, i.e. ferropericlase (x$_{Fe}$ $<$ 0.5) or magnesiumwüstite ($x_{Fe}$ $>$ 0.5). This result is consistent with ferropericlase being a dominant phase not only in the Earth's lower mantle but at the higher pressures of Super-Earth’s mantles. It's worth noting that the lowest enthalpy structure at 200 GPa is the same as the ground-state FeMg$_3$O$_4$ \textit{Cmmm} at 350 GPa shown in Fig.~\ref{fig:fig4}. Inspecting the higher energy range in Fig.~\ref{fig:fig7}(a) and (c), one finds that oxygen-rich structures generally have lower enthalpies than oxygen-poor structures. The oxygen-rich structures mainly contain octahedral clusters, while the oxygen-poor structures can have a greater variety of motifs, including FCC, BCC, HCP-type clusters. This is mainly because Fe and Mg start to alloy to form closely packed motifs under unsaturated oxygen conditions. 

At 350 GPa, structures with 50\% oxygen concentrations still have the lowest formation enthalpy. A few oxygen-rich phases become more stable and approach the convex hull energetically compared to 200 GPa. BCC-type clusters start to appear in these oxygen-rich structures at 350 GPa, indicating that the B2-type Fe-O clusters are favored at higher pressures over  B1-type clusters at lower pressures. The situation with oxygen-poor structures at 350 GPa is similar to the one at 200 GPa. We note that at both 200GPa and 350GPa, several motifs ("other" type) that cannot be classified into the current simple cluster templates appear. Some of them are due to distortions, while some indeed form more complex clusters, e.g., the ones in Fig.~\ref{fig:fig2} and Fig.~\ref{fig:fig3}.

\section{Geophysical implications}
\begin{figure}[hb]
\includegraphics[width=0.48\textwidth]{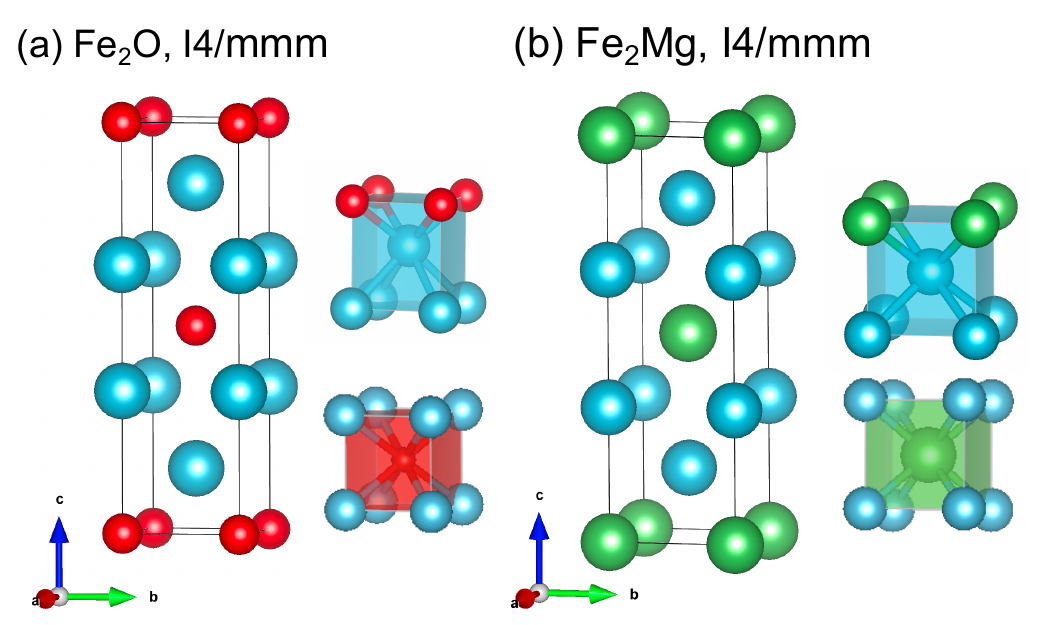}
\caption{\label{fig:fig8} Similar BCC-like Fe$_2$O and Fe$_2$Mg ground states at 350GPa.}
\end{figure}

Our findings on the Fe-Mg-O system at core pressures appear to have some straightforward geophysical consequences. The Fe-rich side (right corner) of the ternary phase diagram in Fig.~\ref{fig:fig1} suggests that Fe$_2$Mg and Fe$_2$O can form a continuous isomorphic solid Fe$_2$(Mg$_{1-x}$O$_x$) solution. Both end-members are BCC-like structures at 350 GPa, as shown in Fig.~\ref{fig:fig8}. BCC-like Fe$_2$Mg and hcp $\epsilon$-Fe are likely to inter-alloy and form a eutectic system, with two coexisting solid phases for some composition-temperature ranges. Small Mg concentrations might produce hcp-like Fe$_{1-x}$Mg$_x$ alloys, but a BCC-like Fe$_{2+x}$Mg$_{1-x}$ might precipitate and coexist beyond a certain concentration threshold. The situation is very similar for the Fe$_2$O-Fe system. Therefore, the Fe-Mg-O system might contain Mg and O dissolved substitutionally in $\epsilon$-Fe for small Mg and O concentrations, but beyond a certain concentration threshold BCC-like Fe$_{2+x+y}$(Mg$_{1/2-x}$O$_{1/2-y}$) might precipitate. BCC-Fe can be stabilized at inner core pressures by alloying with S \cite{52,53}, and it has been argued, but not confirmed, that BCC iron could be stabilized at inner core conditions \cite{54}. Therefore, the precipitation of BCC-like Fe$_{2+x+y}$(Mg$_{1/2-x}$O$_{1/2-y}$) for non-negligible amounts of Mg, O, or both is not a surprising conclusion.

The ternary phases discovered in the O-rich side (left corner) of the phase diagram are relevant for the mantle of some Super-Earths. The absence of stable ternary phases at pressures lower than ~228 GPa suggests that stable phases involving all three elements are solid-solutions of end-member phases with a small concentration of inter-alloying metals. For example, Fig.~\ref{fig:fig7}(a) shows that at 200 GPa, the low-energy structures are dominated by structures with octahedral coordination, with more Mg than Fe, and approximately 50\% O, i.e., ferropericlase or B1-type (Mg$_{1-x}$Fe$_x$)O. At 350 GPa, the oxygen-rich ternary phases FeMgO$_3$, Fe$_2$MgO$_4$, FeMg$_2$O$_4$, and FeMg$_3$O$_4$ emerge as ground states or low-enthalpy phases, besides the B1-type phase. One of them, \textit{I-42d} FeMg$_2$O$_4$, has the same structure as \textit{I-42d} Mg$_2$SiO$_4$, the stable silicate phase predicted to exist in the mantle of Super-Earths above ~500 GPa \cite{17,43}. Here emerges the possibility of an \textit{I-42d}-type Mg$_2$(Si$_{1-x}$Fe$_x$)O$_4$ phase, with Fe substitutional in the Si site, or vice-versa, an unusual type of substitution in the Earth’s mantle, unless as a coupled Mg-Si substitution. From the chemistry standpoint, the newly found phases at 350 GPa can all be viewed as combinations of binary end-members, e.g., FeMgO$_3$ as (MgO)(FeO$_2$), Fe$_2$MgO$_4$ as (MgO)(Fe$_2$O$_3$), FeMg$_2$O$_4$ as (MgO)$_2$(FeO$_2$), and FeMg$_3$O$_4$ as (MgO)$_3$(FeO). Such stable compositions suggest other stable stoichiometric phases might be found by exploring combinations of such end-member compounds, as seen in the Mg-Si-O system, i.e., (MgO)$_n$(SiO$_2$)$_m$ phases \cite{17,43}. Further AGA searches aiming at these complex compositions are needed to identify other possible ternary phases in the Mg-Fe-O system. Finally, O's greater intermixing with the metallic elements at 350 GPa suggests that Mg and O abundances might be non-negligible in the Earth’s inner core. Also, core formation by Fe exsolution from the oxides might be a more complicated process during Super-Earths’ core formation, or O and Mg might be more abundant light elements in Super-Earths’ cores.

\section{Conclusion}
We use the AGA combined with \textit{ab initio} calculations to identify high-pressure structures in the Fe-Mg-O system at 0 K across a wide range of stoichiometries. This procedure is a crucial preparatory stage for modeling the system at finite temperatures. At 350 GPa, we identify mechanically stable phases with FeMg$_2$O$_4$, Fe$_2$MgO$_4$ and FeMg$_3$O$_4$ compositions and one low-enthalpy phase with FeMgO$_3$ composition. These discoveries lead to the construction of the ternary phase diagram and convex hull at 350 GPa. While we have not found any ground-state stoichiometric ternary compound at 200 GPa, the metastable phases' analysis indicates that ferropericlase- or magnesiumwüstite-type phases with 50\% oxygen are very close to the convex hull. Oxygen-rich phases are generally closer to the convex hull than the oxygen-poor phases at all pressures. Motif analyses show octahedral clusters are energetically favored at both pressures and BCC-type clusters start to appear in oxygen-rich phases at 350 GPa. In particular, the nature of iron-rich phases at 350 GPa indicates that Mg, O, or both simultaneously could stabilize a BCC-type iron alloy at inner-core pressures.

\section{acknowledgments}
This work was supported primarily by National Science Foundation awards EAR-1918134 and EAR-1918126 and the Extreme Science and Engineering Discovery Environment (XSEDE), which is supported by National Science Foundation grant number ACI-1548562. R.M.W. and Y.S. also acknowledge partial support from the Department of Energy Theoretical Chemistry Program through grant DOE-DESC0019759. R.W. and Z.L. were supported by the National Natural Science Foundation of China (11774324 \& 12074362) and the Supercomputing Center of USTC. Y.F., F.Z., and S.W. were supported by the National Natural Science Foundation of China (11874307).

\renewcommand{\bibnumfmt}[1]{[#1]}
\bibliographystyle{apsrev4-2}
\end{document}